\documentclass[aps,prb,twocolumn,showpacs]{revtex4}

\usepackage{graphicx}
\usepackage{amsmath}
\usepackage{amssymb}

\begin{document}

\title{Doping dependence of electromagnetic response in cuprate superconductors}

\author{Yiqun Liu, Yingping Mou, and Shiping Feng}
\email{spfeng@bnu.edu.cn}

\affiliation{Department of Physics, Beijing Normal University, Beijing 100875, China}

\begin{abstract}
The study of the electromagnetic response in cuprate superconductors plays a crucial role in the understanding of the essential physics of these materials. Here the doping dependence of the electromagnetic response in cuprate superconductors is studied within the kinetic-energy driven superconducting mechanism. The kernel of the response function is evaluated based on the linear response approximation for a purely transverse vector potential, and can be broken up into its diamagnetic and paramagnetic parts. In particular, this paramagnetic part exactly cancels the corresponding diamagnetic part in the normal-state, and then the Meissner effect is obtained within the entire superconducting phase. Following this kernel of the response function, the electromagnetic response calculation in terms of the specular reflection model qualitatively reproduces many of the striking features observed in the experiments. In particular, the local magnetic-field profile follows an exponential law, while the superfluid density exhibits the nonlinear temperature behavior at the lowest temperatures, followed by the linear temperature dependence extending over the most of the superconducting temperature range. Moreover, the maximal value of the superfluid density occurs at around the {\it critical doping} $\delta_{\rm critical}\sim 0.16$, and then decreases in both lower doped and higher doped regimes. The theory also shows that the nonlinear temperature dependence of the superfluid density at the lowest temperatures can be attributed to the nonlocal effects induced by the d-wave gap nodes on the electron Fermi surface.
\end{abstract}


\maketitle

\section{Introduction}\label{Introduction}

The spontaneous screening of an applied magnetic-field, the Meissner effect, is a defining feature of superconductors \cite{Schrieffer83}, and the detailed screening response as a function of field, temperature, and other parameters is a crucial characterization of cuprate superconductors \cite{Bonn96}. The parent compounds of cuprate superconductors are Mott insulators with an antiferromagnetic (AF) long-range order \cite{Fujita12}, however, with doping a sufficient density of charge carriers into these Mott insulators, the AF long-range ordering weakens and eventually gives to superconductivity leaving the AF short-range order correlation still intact \cite{Fujita12,Bednorz86}. In particular, this superconducting (SC)-state involves pairing of electrons with a d-wave type symmetry \cite{Tsuei00} $\bar{\Delta}({\bf k})\propto\bar{\Delta}(\cos k_{x}-\cos k_{y})$. In this d-wave SC gap, the characteristic feature is the existence of four nodes on the electron Fermi surface (EFS), where the SC gap vanishes. As a nature consequence of doped Mott insulators, the electromagnetic response and its evolution with doping therefore can provide the key information of the correlation between the SC transition temperature $T_{\rm c}$ and the superfluid density $\rho_{\rm s}$ \cite{Bonn96,Uemura89}. On the other hand, since there is a coexistence of the AF short-range order and superconductivity within the entire SC phase \cite{Fujita12}, the magnetic-field can be also used to probe the momentum dependence of the SC gap and spin structure of the electron pair. This is why the first evidence of the SC-state with a d-wave symmetry in cuprate superconductors was obtained from the earlier experimental measurement for $\rho_{\rm s}$ \cite{Hardy93}.

Experimentally, a large body of data available from a wide variety of measurement techniques have provided rather detailed information of the electromagnetic response in cuprate superconductors, where some essential agreements have emerged: (i) the magnetic-field profile follow an exponential field decay \cite{Khasanov04,Suter04}; (ii) $\rho_{\rm s}$ is a linear temperature dependence at the low temperatures except for at the lowest temperatures where a strong deviation from the linear characteristics emerges \cite{Bozovic16,Brewer15,Deepwell13,Khasanov09,Broun07,Panagopoulos99}; (iii) $\rho_{\rm s}$ is strongly dependent on doping, and exhibits a close correlation between $T_{\rm c}$ and $\rho_{\rm s}$ \cite{Uemura89}. In particular, $\rho_{\rm s}$ increases with the increase of doping in the lower doped regime, and reaches a maximum at around the critical doping $\delta\approx 0.19$, then decreases with the increase of doping in the higher doped regime \cite{Lemberger11,Bernhard01}. This special doping dependence of $\rho_{\rm s}$ in turn gives rise to the domelike shape of the doping dependence of $T_{\rm c}$. Although the doping dependence of the electromagnetic response is well-established experimentally \cite{Khasanov04,Suter04,Bozovic16,Brewer15,Deepwell13,Khasanov09,Broun07,Panagopoulos99,Lemberger11,Bernhard01}, its full understanding is still a challenging issue. Theoretically, some physical properties of the electromagnetic response in cuprate superconductors has been extensively studied within the phenomenological Bardeen-Cooper-Schrieffer formalism with a d-wave SC gap \cite{Hone16,Sharapov06,Sheehy04,Kosztin97,Yip92}. In particular, in the local limit, where the magnetic-field penetration depth $\lambda$ is much larger than the coherence length $\zeta$, i.e., $\lambda\gg\zeta$, it has been demonstrated \cite{Kosztin97,Yip92} that the line nodes in a d-wave SC gap produce a linear temperature dependence of the magnetic-field penetration depth $\lambda(T)\propto T/\bar{\Delta}$ at low temperatures, where $\bar{\Delta}$ is the d-wave gap amplitude at zero temperature. In fact, this linear temperature dependence of $\lambda(T)$ at the low temperatures is the hallmark of a clean-limit d-wave superconductor \cite{Hardy93,Jackson00,Kamal98,Lee96}, and in turn leads naturally to a linear temperature dependence of $\rho_{\rm s}(T)$. On the other hand, it has been also shown that the d-wave gap nodes on EFS induces a nonlinear effect of the field on $\lambda(T)$ at the lowest temperatures, associated with a field induced an increase in the density of the quasiparticle excitation states located at around the d-wave gap nodes on EFS \cite{Kosztin97,Yip92}. This follows from a fact that the nonlocal effect is closely related to the divergence of the coherence length $\zeta$ at the d-wave gap nodes on EFS, as the coherence length $\zeta$ varies in inverse proportion to the d-wave SC gap amplitude, then this nonlocal effect at the lowest temperatures can induce a nonlinear temperature dependence of $\lambda(T)$ in the clean-limit.

In the early studies \cite{Feng10}, the electromagnetic response in cuprate superconductors has been discussed within the kinetic-energy driven SC mechanism \cite{Feng0306,Feng12,Feng15}, where in the {\it decoupling approximation}, the electron polarization operator and the related electron current density operator are identified approximately to the corresponding charge-carrier polarization operator and charge-carrier current density operator \cite{Feng10}. However, an obvious insufficiency in this {\it decoupling approximation} is that the charge-carrier Green's function does not produces a large EFS observed from the angle-resolved photoemission spectroscopy (ARPES) experiments. Therefore these early studies of the electromagnetic response is rather incomplete. Following the kinetic-energy driven superconductivity \cite{Feng0306,Feng12,Feng15}, we \cite{Feng15a} have developed recently a full charge-spin recombination scheme to fully recombine a charge carrier and a localized spin into an electron, where the obtained electron Green's function can produce a large EFS with the area that fulfills Luttinger's theorem. As a complement of the previous analysis of the electromagnetic response in cuprate superconductors \cite{Feng10}, we in this paper restudy the doping dependence of the electromagnetic response in cuprate superconductors within the kinetic-energy driven superconductivity, where we employ the {\it electron Green's function} to calculate the kernel of the response function based on the linear response approximation for a purely transverse vector potential. In particular, this kernel of the response function can be broken up into its diamagnetic and paramagnetic parts. However, the paramagnetic part of the kernel of the response function exactly cancels the corresponding diamagnetic part of the kernel of the response function in the normal-state, and then the Meissner effect is obtained within the entire SC phase. Following this kernel of the response function, the electromagnetic response calculation in terms of the specular reflection model qualitatively reproduces the main features observed in the experiments \cite{Khasanov04,Suter04,Bozovic16,Brewer15,Deepwell13,Khasanov09,Broun07,Panagopoulos99,Lemberger11,Bernhard01,Jackson00,Kamal98,Lee96}. In particular, the local magnetic-field profile follows an exponential law, while $\rho_{\rm s}(T)$ exhibits the nonlinear temperature behavior at the lowest temperatures, followed by the linear temperature dependence extending over the most of the SC temperature range. Our theory also indicates that the nonlinear temperature dependence of $\rho_{\rm s}(T)$ at the lowest temperatures is induced by the nonlocal effect due to the presence of the d-wave gap nodes on EFS.

This paper is organized as follows. The general framework of the electromagnetic response within the kinetic-energy driven superconductivity is presented in Section \ref{general-framework}. In Section \ref{meissner-effect}, we study the characteristic features of the electromagnetic response in a weak electromagnetic field, where $\rho_{\rm s}$ exhibits a maximum value at around the critical doping $\delta\approx 0.16$, and then decreases at both lower doped and higher doped regimes. Finally, we give a summary and discussions in Section \ref{conclusions}.

\section{General framework of electromagnetic response}\label{general-framework}

\subsection{Linear response theory}\label{Linear-response}

For the discussions of the evolution of the electromagnetic response with doping in cuprate superconductors, we start from the general relation between the electron current density ${\bf J}$ and vector potential ${\bf A}$ \cite{Fukuyama69,Misawa94,Kostyrko94}:
\begin{equation}\label{linres}
J_\mu({\bf q},\omega)=-\sum\limits_{\nu=1}^{3} K_{\mu\nu}({\bf q},\omega)A_{\nu}({\bf q},\omega),
\end{equation}
where the Greek indices label the axes of the Cartesian coordinate system, while $K_{\mu\nu}$ is a nonlocal kernel of the response function. In particular, this kernel of the response function (\ref{linres}) can be broken up into its diamagnetic (d) and paramagnetic (p) parts as,
\begin{equation}\label{kernel}
K_{\mu\nu}({\bf q},\omega)=K^{({\rm d})}_{\mu\nu}({\bf q},\omega)+K^{({\rm p})}_{\mu\nu}({\bf q},\omega),
\end{equation}
which is closely associated with the electron current density in the presence of the vector potential ${\bf A}$, and therefore plays a crucial role for the understanding of the doping dependence of the electromagnetic response in cuprate superconductors.

\subsection{Electron Green's function}\label{EGF}

In cuprate superconductors, the single common feature is the presence of the two-dimensional copper-oxide layers \cite{Kastner98}, and then it is believed that the anomalous properties of cuprate superconductors are closely related to these copper-oxide layers. It is commonly accepted that the essential physics of the copper-oxide layer \cite{Anderson87} can be described by the $t$-$J$ model on a square lattice with the nearest-neighbor (NN) spin-spin AF exchange along with the electron hopping. However, for the discussions of the doping dependence of the electromagnetic response, the $t$-$J$ model should be extended by including the exponential Peierls factor as \cite{Feng10,Fukuyama69,Misawa94,Kostyrko94},
\begin{eqnarray}\label{tjham}
H&=&-\sum_{l\hat{a}\sigma}t_{\hat{a}}e^{-i(e/{\hbar}){\bf A}(l)\cdot\hat{a}}C^{\dagger}_{l\sigma}C_{l+\hat{a}\sigma}+\mu\sum_{l\sigma} C^{\dagger}_{l\sigma}C_{l\sigma}\nonumber\\
&+& J\sum_{l\hat{\eta}}{\bf S}_{l}\cdot {\bf S}_{l+\hat{\eta}},~~~~~~
\end{eqnarray}
where the summation is over all sites $l$, and for each $l$, over its NN sites $\hat{a}=\hat{\eta}$ with the hopping amplitude $t_{\hat{a}}= t_{\hat{\eta}}=t$ or next NN sites $\hat{a}=\hat{\eta}'$ with the hopping amplitude $t_{\hat{a}}=t_{\hat{\eta}'}=-t'$, while the spin-spin interaction occurs only for the NN sites $\hat{\eta}$. $C^{\dagger}_{l\sigma}$ ($C_{l\sigma}$) is the electron creation (annihilation) operator with spin $(\uparrow,\downarrow)$, ${\bf S}_{l}=(S^{x}_{l}, S^{y}_{l}, S^{z}_{l})$ is the spin operator, and $\mu$ is the chemical potential. In the $t$-$J$ model (\ref{tjham}), there is a local constraint of no double electron occupancy $\sum_{\sigma}C^{\dagger}_{l\sigma} C_{l\sigma}\leq 1$, while the exponential Peierls factor accounts for the coupling of electrons to an external magnetic field in terms of the vector potential ${\bf A}(l)$. The strong electron correlation in the $t$-$J$ model manifests itself by the electron single occupancy local constraint \cite{Anderson87}, which can be treated properly in analytical calculations within the charge-spin separation (CSS) fermion-spin theory \cite{Feng15,Feng9404}, where the constrained electron operators in the $t$-$J$ model (\ref{tjham}) are decoupled as $C_{l\uparrow}=h^{\dagger}_{l\uparrow}S^{-}_{l}$ and $C_{l\downarrow}=h^{\dagger}_{l\downarrow}S^{+}_{l}$, with the spinful fermion operator $h_{l\sigma}=e^{-i\Phi_{i\sigma}}h_{l}$ describes the charge degree of freedom together with some effects of spin configuration rearrangements due to the presence of the doped charge carrier itself, while the spin operator $S_{l}$ describes the spin degree of freedom, then the electron local constraint of no double electron occupancy is satisfied in analytical calculations. In this CSS fermion-spin representation, the $t$-$J$ model (\ref{tjham}) can be rewritten as \cite{Feng15,Feng9404},
\begin{eqnarray}\label{cssham}
H&=&\sum_{l\hat{a}}t_{\hat{a}}e^{-i{e\over \hbar}{\bf A}(l)\cdot\hat{a}}(h^{\dagger}_{l+\hat{a}\uparrow}h_{l\uparrow}S^{+}_{l}S^{-}_{l+\hat{a}} +h^{\dagger}_{l+\hat{a}\downarrow}h_{l\downarrow}S^{-}_{l}S^{+}_{l+\hat{a}})\nonumber\\
&-&\mu\sum_{l\sigma}h^{\dagger}_{l\sigma}h_{l\sigma}+J_{{\rm eff}}\sum_{l\hat{\eta}}{\bf S}_{l}\cdot {\bf S}_{l+\hat{\eta}},
\end{eqnarray}
with $J_{{\rm eff}}=(1-\delta)^{2}J$, and $\delta=\langle h^{\dagger}_{l\sigma}h_{l\sigma}\rangle=\langle h^{\dagger}_{l}h_{l}\rangle$ is the doping concentration. In the following discussions, the parameters in the $t$-$J$ model are chosen as $t/J=3.4$ and $t'/J=1.2$. However, when necessary to compare with the experimental data, we take $J=1000$ K, which is the typical value of cuprate superconductors.

The electron pairs are crucial for superconductivity because these electron pairs behave as effective bosons, and can form something analogous to a Bose condensate that flows without resistance. In conventional superconductors, as explained by the electron-phonon mechanism \cite{Schrieffer83,Bardeen57,Eliashberg60,McMillan65}, the interaction between electrons by the exchange of phonons drives the formation of the electron pairs responsible for superconductivity. However, what type of the collective bosonic excitation that acts like a bosonic glue to hold the electron pairs together in cuprate superconductors still is disputed. Based on the CSS fermion-spin formulation of the $t$-$J$ model (\ref{cssham}), we \cite{Feng0306,Feng12,Feng15} have established a kinetic-energy driven SC mechanism, where the interaction between charge carriers and spins directly from the kinetic energy by the exchange of spin excitations generates the formation of the d-wave charge-carrier pairs, while the d-wave electron pairs originate from the d-wave charge-carrier pairing state are due to the charge-spin recombination, and their condense into the d-wave SC-state. In cuprate superconductors, EFS plays a crucial role in the understanding of the anomalous properties, since everything of the low-energy behavior happens at EFS. However, in the conventional charge-spin recombination scheme, EFS observed from the ARPES experiments can not be restored \cite{Feng93}. Recently, a full charge-spin recombination scheme has been developed within the kinetic-energy driven SC mechanism to incorporate EFS \cite{Feng15a}, where a charge carrier and a localized spin are fully recombined into a physical electron, and then a large EFS is reproduced with the area that fulfills Luttinger's theorem. Within this full charge-spin recombination scheme \cite{Feng15a}, the full electron normal and anomalous Green's functions of the $t$-$J$ model (\ref{cssham}) have been evaluated, where the electron self-energy in the particle-particle channel is related directly to the SC gap, while the electron self-energy in the particle-hole channel is closely associated with the single-particle coherence. In particular, these full electron normal and anomalous Green's functions have been employed to discuss the electronic state properties of cuprate superconductors \cite{Gao19,Feng16,Gao18}, and the obtained results are well consistent with the experimental observations. In this paper, as a qualitative discussion of the doping dependence of the electromagnetic response in cuprate superconductors, the single-particle coherence can be generally studied in the static-limit approximation. In this case, the full electron normal and anomalous Green's functions in the zero magnetic field case have been obtained explicitly in the Nambu representation \cite{Feng15a},
\begin{equation}\label{NPEGF}
\mathbb{G}({\bf k},\omega) = Z_{\rm F}{\omega\tau_{0} + \bar{\varepsilon}_{\bf k}\tau_{3}-\bar{\Delta}_{\rm Z}({\bf k})\tau_{1}\over\omega^{2}- E_{\bf k}^{2}},
\end{equation}
where $\tau_{0}$ is a unit matrix, $\tau_{1}$ and $\tau_{3}$ are Pauli matrices, $\bar{\varepsilon}_{\bf k}=Z_{\rm F}\varepsilon_{\bf k}$, with the bare electron dispersion $\varepsilon_{\bf k}=-Zt\gamma_{\bf k}+Zt'\gamma_{\bf k}'+\mu$, $\gamma_{\bf k}=({\rm cos}k_{x}+{\rm cos}k_{y})/2$, $\gamma_{\bf k}'= {\rm cos} k_{x}{\rm cos}k_{y}$, and the number of NN or the next NN sites on a square lattice $Z$, $\bar{\Delta}_{\rm Z}({\bf k})= Z_{\rm F}\bar{\Delta}({\bf k})$, with the d-wave SC gap $\bar{\Delta}({\bf k})=\Sigma_{2}({\bf k},\omega=0)=\bar{\Delta}({\rm cos} k_{x}-{\rm cos} k_{y})/2$, $E_{\bf k}=\sqrt{\bar{\varepsilon}^{2}_{{\bf k}}+|\bar{\Delta}_{\rm Z}({\bf k})|^{2}}$ is the SC quasiparticle spectrum, $Z^{-1}_{\rm F} =1-{\rm Re}\Sigma_{\rm 1o}({\bf k},\omega=0) \mid_{{\bf k}=[\pi,0]}$ is single-particle coherent weight, with the antisymmetric part $\Sigma_{\rm 1o}({\bf k},\omega)$ of the electron self-energy $\Sigma_{1}({\bf k},\omega)$, while the electron self-energies $\Sigma_{1}({\bf k},\omega)$ in the particle-hole channel and $\Sigma_{2}({\bf k},\omega)$ in the particle-particle channel have been given explicitly in Ref. \onlinecite{Feng15a}. Moreover, these SC gap parameter $\bar{\Delta}$, the single-particle coherent weight $Z_{\rm F}$, and the chemical potential $\mu$ have been determined self-consistently without using any adjustable parameters \cite{Feng15a}.

\subsection{Kernel of response function} \label{kernel-function}

In the linear response relation (\ref{linres}), the vector potential ${\bf A}$ is coupled to the electrons, which are now represented by the electron operators $C_{l\uparrow}= h^{\dagger}_{l\uparrow}S^{-}_{l}$ and $C_{l\downarrow}=h^{\dagger}_{l\downarrow}S^{+}_{l}$ in the CSS fermion-spin representation \cite{Feng9404,Feng15}. For the calculation of the electron current density, we firstly need to obtain the electron polarization operator, which is defined as a summation over all the particles and their positions, and can be obtained directly in the present CSS fermion-spin representation as,
\begin{equation}\label{poloper}
{\bf P}=-e\sum\limits_{l\sigma}{\bf R}_{l}C^{\dagger}_{l\sigma}C_{l\sigma}=e\sum\limits_{l}{\bf R}_{l}h^{\dagger}_{l} h_{l},
\end{equation}
then the electron current density operator is obtained explicitly by evaluating the time-derivative of the above polarization operator (\ref{poloper}) as \cite{Feng10,Mahan81},
\begin{eqnarray}\label{current-density-operator}
{\bf J}&=&{\partial {\bf P}\over\partial t}={i\over\hbar}[H,{\bf P}]\nonumber\\
&=&{ie\over\hbar}\sum\limits_{l{\hat{a}}}t_{\hat{a}}{\hat{a}}e^{-i{e\over\hbar}{\bf A}(l)\cdot{\hat{a}}}(h_{l\uparrow} h^{\dagger}_{l+\hat{a}\uparrow}S^{+}_{l}S^{-}_{l+\hat{a}}\nonumber\\
&+&h_{l\downarrow}h^{\dagger}_{l+\hat{a}\downarrow}S^{-}_{l}S^{+}_{l+\hat{a}}) \nonumber\\
&=&{ie\over\hbar}\sum\limits_{l{\hat{a}}\sigma}t_{\hat{a}}{\hat{a}}e^{-i{e\over\hbar}{\bf A}(l)\cdot{\hat{a}}}C^{\dagger}_{l\sigma} C_{l+\hat{a}\sigma}.~~~~~
\end{eqnarray}
In the corresponding to the diamagnetic and paramagnetic parts of the kernel of the response function in Eq. (\ref{kernel}),
this electron current density operator in Eq. (\ref{current-density-operator}) also can be break up into its diamagnetic (d) and paramagnetic (p) parts as ${\bf J}={\bf J}^{({\rm d})} + {\bf j}^{({\rm p})}$, with the diamagnetic and paramagnetic parts of the electron current density operator that can be expressed in the linear response approximation as,
\begin{subequations}
\begin{eqnarray}
{\bf J}^{({\rm d})}&=&{e^{2}\over\hbar^{2}}\sum\limits_{l{\hat{a}}\sigma}t_{\hat{a}}{\hat{a}}{\bf A}(l)\cdot{\hat{a}}C^{\dagger}_{l\sigma} C_{l+\hat{a}\sigma},\label{tcurdia}\\
{\bf J}^{({\rm p})}&=&{ie\over\hbar}\sum\limits_{l{\hat{a}}\sigma}t_{\hat{a}}{\hat{a}}C^{\dagger}_{l\sigma}C_{l+\hat{a}\sigma}, \label{tcurpara9}
\end{eqnarray}
\end{subequations}
respectively, where the diamagnetic part of the electron current density operator is proportional to the vector potential, and then the diamagnetic part of the response kernel can be derived directly as,
\begin{eqnarray}\label{diakernel}
K_{\mu\nu}^{({\rm d})}({\bf q},\omega)={4e^{2}\over\hbar^{2}}(\phi_{\rm c1}t-2\phi_{\rm c2}t')\delta_{\mu\nu}={1\over\lambda^{2}_{L}} \delta_{\mu\nu},
\end{eqnarray}
where $\lambda_{L}$ is so-called London penetration depth, while the electron particle-hole parameters $\phi_{\rm c1}=\langle C^{\dagger}_{l\sigma} C_{l+\hat{\eta}\sigma}\rangle$ and $\phi_{\rm c2}=\langle C^{\dagger}_{l\sigma}C_{l+\hat{\eta}'\sigma}\rangle$ are calculated from the electron normal Green's function in Eq. (\ref{NPEGF}) as,
\begin{subequations}\label{SCE2}
\begin{eqnarray}
\phi_{\rm c1}&=&{1\over 2N}\sum_{{\bf k}}\gamma_{{\bf k}}Z_{\rm F}\left (1-{\bar{\varepsilon}_{{\bf k}}\over E_{\bf k}}{\rm tanh}[{1\over 2}\beta E_{\bf k}]\right ),~~~~\\
\phi_{\rm c2}&=&{1\over 2N}\sum_{{\bf k}}\gamma_{{\bf k}}'Z_{\rm F}\left (1-{\bar{\varepsilon}_{{\bf k}}\over E_{\bf k}}{\rm tanh}[{1\over 2}\beta E_{\bf k}]\right ),~~~
\end{eqnarray}
\end{subequations}
with the number of sites on a square lattice $N$. Since these $\phi_{\rm c1}$ and $\phi_{\rm c2}$ are doping and temperature dependent, leading to that $\lambda_{L}$ is also doping and temperature dependent.

The paramagnetic part of the response kernel, on the other hand, can be obtained as $K_{\mu\nu}^{({\rm p})}({\bf q},\omega) =P_{\mu\nu}({\bf q}, \omega)$, with the electron current-current correlation function \cite{Feng10,Fukuyama69,Misawa94,Kostyrko94},
\begin{equation}\label{corP}
P_{\mu\nu}({\bf q},\tau)=-\langle T_{\tau}J^{({\rm p})}_{\mu}({\bf q},\tau)J_{\nu}^{({\rm p})}(-{\bf q},0)\rangle .
\end{equation}
For the calculation of the above electron current-current correlation function (\ref{corP}), it is convenient to work in the Nambu representation, where the electron Nambu operators are defined as $\Psi^{\dagger}_{\bf k}=(C^{\dagger}_{{\bf k}\uparrow},C_{-{\bf k}\downarrow})$ and $\Psi_{{\bf k}+{\bf q}} =(C_{{\bf k}+{\bf q}\uparrow},C^{\dagger}_{-{\bf k}-{\bf q}\downarrow})^{\rm T}$.

The electron density operator is summed over the position of all electrons, and then its Fourier transform can be expressed as $\rho({\bf q})= (e/N) \sum_{{\bf k}\sigma} C^{\dagger}_{{\bf k}\sigma} C_{{\bf k}+{\bf q} \sigma} = (e/N)\sum_{{\bf k}} \Psi^{\dagger}_{\bf k} \tau_{3} \Psi_{{\bf k}+{\bf q}}$. In this Nambu representation, the paramagnetic four-current density operator can be represented as,
\begin{eqnarray}\label{curnambu}
J_{\mu}^{({\rm p})}({\bf q})={1\over N}\sum\limits_{\bf k}\Psi^{\dagger}_{{\bf k}}{\mathbf{\gamma}}_{\mu}({\bf k},{\bf q})\Psi_{{\bf k}+{\bf q}}.
\end{eqnarray}
with the bare current vertex,
\begin{eqnarray}
{\mathbf{\gamma}}_{\mu}({\bf k},{\bf q})= \left
\{\begin{array}{ll} -{2e\over\hbar}\, e^{{1\over 2}iq_{\mu}}
\{\sin(k_{\mu}+{1\over 2}q_{\mu})\\
\times [t-2t'\sum\limits_{\nu\neq\mu}\cos({1\over 2}q_{\nu})\cos(
k_{\nu}+{1\over 2}q_{\nu})]\\
-i(2t')\cos(k_{\mu}+{1\over 2}q_{\mu})\\
\times \sum\limits_{\nu\neq\mu}\sin ({1\over 2}q_{\nu})\sin(k_{\nu}+{1\over 2}q_{\nu})
\}\tau_{0} ~~ {\rm for}\ \mu\neq 0,\\
e\tau_{3} ~~ {\rm for}\ \mu=0.\\
\end{array}\right.\label{barevertex}
\end{eqnarray}
As in our previous discussions \cite{Feng10}, we are calculating the electron current-current correlation function (\ref{corP}) with the paramagnetic current density operator (\ref{curnambu}), i.e., bare current vertex (\ref{barevertex}), but Green function (\ref{NPEGF}). As a consequence, we do not take into account longitudinal excitations properly \cite{Schrieffer83,Misawa94}, the obtained results are valid only in the gauge, where the vector potential is purely transverse, e.g. in the Coulomb gauge. In this case, the electron current-current correlation function (\ref{corP}) can be derived in the Nambu representation as,
\begin{eqnarray}
P_{\mu\nu}(&{\bf q},&i\omega_{n})={1\over N}\sum\limits_{\bf k}{\mathbf{\gamma}}_{\mu}({\bf k},{\bf q}){\mathbf{\gamma}}^{*}_{\nu}({\bf k},{\bf q}) \nonumber\\
&\times& {1\over\beta}\sum\limits_{i\omega_{m}}{\rm{Tr}}\left[{\mathbb{G}}({\bf k}+{\bf q},i\omega_{n}+i\omega_{m}){\mathbb{G}}({\bf k}, i\omega_{m})\right]. ~~~~~~~\label{barepolmats}
\end{eqnarray}
With the help of the electron Green's function in Eq. (\ref{NPEGF}), the paramagnetic part of the response kernel $K_{\mu\nu}^{({\rm p})}({\bf q}, \omega) =P_{\mu\nu}({\bf q},\omega)$ in the static limit ($\omega\sim 0$) can be obtained as,
\begin{eqnarray}\label{parakernel}
K_{\mu\nu}^{({\rm p})}({\bf q},0)&=&{1\over N}\sum_{\bf k}\mathbf{\gamma}_{\mu}({\bf k},{\bf q})\mathbf{\gamma}_{\nu}^*({\bf k},{\bf q})
[L_{\rm c1}({\bf k},{\bf q})+L_{\rm c2}({\bf k},{\bf q})]\nonumber\\
&=&K_{\mu\mu}^{({\rm p})}({\bf q},0)\delta_{\mu\nu},~~~~
\end{eqnarray}
where the functions $L_{\rm c1}({\bf k},{\bf q})$ and $L_{\rm c2}({\bf k},{\bf q})$ are given explicitly by,
\begin{subequations}
\begin{eqnarray}
L_{\rm c1}({\bf k},{\bf q})&=&Z_{\rm F}^{2}\left [1+{\bar{\varepsilon}_{{\bf k}+{\bf q}}\bar{\varepsilon}_{\bf k}+\bar{\Delta}_{\rm Z}({\bf k}+{\bf q}) \bar{\Delta}_{\rm Z}({\bf k})\over E_{\bf k}E_{{\bf k}+{\bf q}}}\right ]\nonumber\\
&\times& {n_{\rm F}(E_{\bf k})-n_{\rm F}(E_{{\bf k}+{\bf q}})\over E_{\bf k}-E_{{\bf k} +{\bf q}}}, ~~~\\
L_{\rm c2}({\bf k},{\bf q})&=&Z_{\rm F}^{2}\left [1-{\bar{\varepsilon}_{{\bf k}+{\bf q}}\bar{\varepsilon}_{\bf k}+\bar{\Delta}_{\rm Z}({\bf k}+{\bf q}) \bar{\Delta}_{\rm Z}({\bf k})\over E_{\bf k}E_{{\bf k}+{\bf q}}}\right ]\nonumber\\
&\times& {n_{\rm F}(E_{\bf k})+n_{\rm F}(E_{{\bf k}+{\bf q}})-1\over E_{\bf k}+E_{{\bf k}+{\bf q}}},~~~~~
\end{eqnarray}
\end{subequations}
Now we can obtain the kernel of the response function in Eq. (\ref{kernel}) from Eqs. (\ref{diakernel}) and (\ref{parakernel}) as,
\begin{eqnarray}
K_{\mu\nu}({\bf q},0)=\left [{1\over\lambda^{2}_{\rm L}}+K_{\mu\mu}^{({\rm p})}({\bf q},0)\right ]\delta_{\mu\nu}. \label{kernel1}
\end{eqnarray}

\subsection{Superfluid density in the long wavelength limit} \label{long-wavelength-limit}

In the long wavelength limit $|{\bf q}|\to 0$, the function $L_{\rm c2}({\bf k},{\bf q}\to 0)$ is equal to zero, and then the paramagnetic part of the response kernel in Eq. (\ref{parakernel}) is reduced as,
\begin{eqnarray}
K_{yy}^{({\rm p})}({\bf q}\to 0,0)&=& 2Z^{2}_{\rm{F}}{4e^{2}\over\hbar^{2}}{1\over N}\sum\limits_{\bf k}\sin^{2} k_{y}[t-2t'\cos k_{x}]^{2} \nonumber\\
&\times& \lim\limits_{{\bf q}\to 0}{n_{\rm F}(E_{\bf k})-n_{\rm F}(E_{{\bf k}+{\bf q}})\over E_{\bf k}-E_{{\bf k}+{\bf q}}}.~~~~~~ \label{kernel5}
\end{eqnarray}
Firstly, we discuss two extreme cases: (i) at the temperature $T=0$, we find that the paramagnetic part of the response kernel (\ref{kernel5}) is reduced further as,
\begin{eqnarray}
K_{yy}^{({\rm p})}(&{\bf q}&\to 0,0)|_{T\to 0}=-2Z^{2}_{\rm F}{4e^{2}\over\hbar^{2}}{1\over N}\sum\limits_{\bf k}\sin^{2} k_{y}\nonumber\\
&\times&[t- 2t'\cos k_{x}]^{2}{\beta e^{\beta E_{\bf k}}\over (e^{\beta E_{\bf k}}+1)^{2}}|_{T\to 0}. ~~~~\label{kernel6}
\end{eqnarray}
As we have mentioned in the above Sec. \ref{Introduction}, the characteristic feature of the d-wave SC gap is that it vanishes along the diagonal directions of the Brillouin zone (BZ), which leads to that the right-hand side of the above paramagnetic part of the response kernel (\ref{kernel6}) can be separated into two terms:
\begin{eqnarray}
K_{yy}^{({\rm p})}(&{\bf q}&\to 0,0)|_{T\to 0} = -2Z^{2}_{\rm F}{4e^{2}\over\hbar^{2}}{1\over N}\sum\limits_{{\bf k}(|k_{x}|\neq |k_{y}|)}\sin^{2} k_{y} \nonumber\\
&\times& [t-2t'\cos k_{x}]^{2}{\beta e^{\beta E_{\bf k}}\over (e^{\beta E_{\bf k}}+1)^{2}}|_{T\to 0}\nonumber\\
&-&2Z^{2}_{\rm F}{4e^{2}\over\hbar^{2}}{1\over N_{x}}{1\over N_{y}}\sum\limits_{k_{y}}\sin^{2} k_{y}\nonumber\\
&\times& [t-2t'\cos k_{y}]^{2}{\beta e^{\beta \bar{\varepsilon}_{k_{y}}}\over (e^{\beta\bar{\varepsilon}_{k_{y}}}+1)^{2}}|_{T\to 0} ,  \label{kernel7}
\end{eqnarray}
where $N=N_{x}N_{y}$, with $N_{x}$ ($N_{y}$) that is the corresponding number of sites in the $\hat{x}$ ($\hat{y}$) direction, $\bar{\varepsilon}_{k_{y}}=-Z_{\rm F}(Zt\cos k_{y}-Zt'\cos^{2} k_{y}-\mu)$. Since the SC gap $\bar{\Delta}({\bf k})\neq 0$ except for along the diagonal directions of BZ, the first term of the right-hand side in Eq. (\ref{kernel7}) is equal to zero. However, to show that the second term of the right-hand side in Eq. (\ref{kernel7}) is also equal to zero, we rewrite it as,
\begin{eqnarray}
&-&2Z^{2}_{\rm F}{4e^{2}\over\hbar^{2}}{1\over N_{x}}\int^{\pi}_{-\pi}{dk_{y}\over 2\pi}\sin^{2} k_{y}[t-2t'\cos k_{y}]^{2}\nonumber\\
&\times& {\beta e^{\beta\bar{\varepsilon}_{k_{y}}}\over (e^{\beta\bar{\varepsilon}_{k_{y}}}+1)^{2}}|_{T\to 0}\nonumber\\
&=&{2Z_{\rm F}\over Z}{4e^{2}\over\hbar^{2}}{1\over N_{x}}\int^{\pi}_{-\pi}{dk_{y}\over 2\pi}[t\cos k_{y}-2t'\cos (2k_{y})] \nonumber\\
&\times& {1\over e^{\beta\bar{\varepsilon}_{k_{y}}}+1}|_{T\to 0}\nonumber\\
&=&{2Z_{\rm F}\over Z}{4e^{2}\over\hbar^{2}}{1\over N_{x}}\int^{\pi}_{-\pi}{dk_{y}\over 2\pi}[t\cos k_{y}-2t'\cos (2k_{y})] \theta(\bar{\varepsilon}_{k_{y}}),~~~~~
\end{eqnarray}
which is equal to zero in the thermodynamic limit $N\to\infty$ (then $N_{x}\to\infty$ and $N_{y}\to\infty$), where the step function $\theta(x)=1$ for $x<0$ and $\theta(x)=0$ for $x>0$. These results reflect a fact that at $T=0$, the long wavelength electromagnetic response is determined only by the diamagnetic part of the response kernel, i.e., the kernel of the response function in Eq. (\ref{kernel1}) is reduced as, $K_{\mu\nu}({\bf q}\to 0,0)|_{T=0}=\lambda^{-2}_{L}\delta_{\mu\nu}$.

(ii) Now we turn to discuss the extreme case at $T=T_{\rm c}$ ($\beta_{c}=T^{-1}_{\rm c}$), where the SC gap $\bar{\Delta}({\bf k})|_{T=T_{\rm c}} =0$, and then the paramagnetic part of the response kernel in Eq. (\ref{kernel5}) is reduced as,
\begin{eqnarray}
&&K_{yy}^{({\rm p})}({\bf q}\to 0,0)=-2Z^{2}_{\rm F}{4e^{2}\over\hbar^{2}}\int^{\pi}_{-\pi}{dk_{x}\over 2\pi}\int^{\pi}_{-\pi}{dk_{y}\over 2\pi}\sin^{2} k_{y}\nonumber\\
&\times&[t-2t'\cos k_{x}]^{2}{\beta_{c}e^{\beta_{c}\bar{\varepsilon}_{k}}\over (e^{\beta_{c}\bar{\varepsilon}_{k}}+1)^{2}}\nonumber\\
&=&-Z_{\rm F}{4e^{2}\over\hbar^{2}}\int^{\pi}_{-\pi}{dk_{x}\over 2\pi}\int^{\pi}_{-\pi}{dk_{y}\over 2\pi}[t\cos k_{y}-2t'\cos k_{x}\cos k_{y}] \nonumber\\
&\times& {1\over e^{\beta_{c}\bar{\varepsilon}_{k}}+1}\nonumber\\
&=&-Z_{\rm F}{4e^{2}\over\hbar^{2}}{1\over N}\sum\limits_{\bf k}[t\cos k_{y}-2t'\cos k_{x}\cos k_{y}]n_{F}(\bar{\varepsilon}_{k})\nonumber\\
&=&-{4e^{2}\over\hbar^{2}}[\phi_{\rm c1}t-2\phi_{\rm c2}t'] =-{1\over \lambda^{2}_{L}},
\end{eqnarray}
which exactly cancels the diamagnetic part of the response kernel in Eq. (\ref{diakernel}), then the Meissner effect occurs within the entire SC phase.

\begin{figure}[h!]
\centering
\includegraphics[scale=1.10]{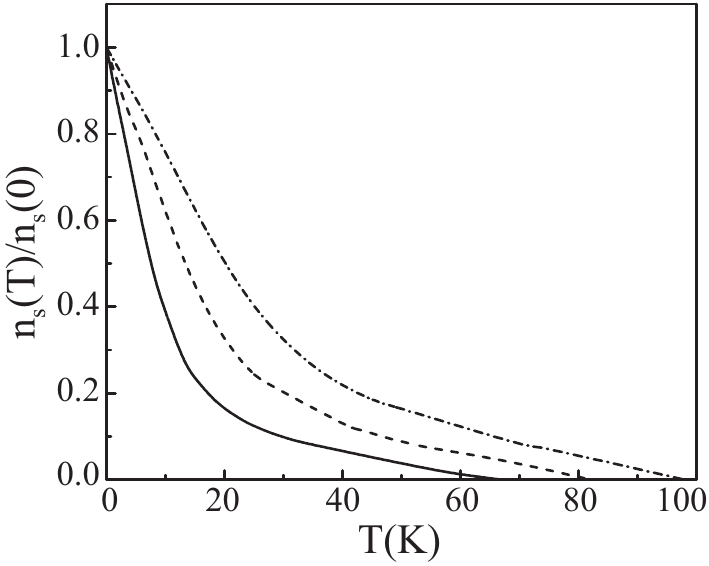}
\caption{The normalized effective-superfluid density as a function of temperature at $\delta=0.08$ (solid line), $\delta=0.12$ (dashed line), and $\delta=0.15$ (dash-dotted line) for $t/J=3.4$, $t'/J=1.2$, and $J=1000$ K. \label{effectivsuperfluid}}
\end{figure}

To show this doping and temperature dependence of Meissner effect more clearly, the paramagnetic part of the response kernel in Eq. (\ref{kernel5}) can be rewritten as,
\begin{eqnarray}
K_{\mu\nu}^{({\rm p})}({\bf q}\to 0,0)=-{1\over\lambda^{2}_{L}}\left [1-{n_{\rm s}(T)\over n_{\rm s}(0)}\right ]\delta_{\mu\nu}, \label{kerne30}
\end{eqnarray}
where the normalized effective-superfluid density is given by,
\begin{eqnarray}
{n_{\rm s}(T)\over n_{\rm s}(0)}&=&1-2\lambda^{2}_{L}Z^{2}_{\rm F}{4e^{2}\over\hbar^{2}}{1\over N}\sum\limits_{\bf k}\sin^{2}k_{y}[t-2t'\cos k_{x}]^{2}\nonumber\\
&\times& {\beta e^{\beta E_{\bf k}}\over (e^{\beta E_{\bf k}}+1)^{2}}. ~~~\label{ratio2}
\end{eqnarray}
Substituting the paramagnetic part of the response kernel (\ref{kerne30}) into Eq. (\ref{kernel1}), the kernel of the response function is expressed explicitly in terms of the effective superfluid density as,
\begin{eqnarray}
K_{\mu\nu}({\bf q}\to 0,0)={1\over\lambda^{2}_{L}}{n_{\rm s}(T)\over n_{\rm s}(0)}\delta_{\mu\nu}. \label{kernel2}
\end{eqnarray}
In Fig. \ref{effectivsuperfluid}, we plot the normalized effective-superfluid density as a function of temperature $T$ for the doping concentration $\delta=0.08$ (solid line), $\delta=0.12$ (dashed line), and $\delta=0.15$ (dash-dotted line), where the most typical features can be summarized as: (i) the effective superfluid density decreases with the increase of temperatures, and disappears at $T_{\rm c}$, indicating that the Meissner effect in cuprate superconductors occurs within the entire SC phase, while all the electron quasiparticles are in the normal fluid in the normal-state ($T> T_{\rm c}$); (ii) the electromagnetic response kernel has a London form in the long wavelength limit. In particular, it should be emphasized that although the electromagnetic response kernel is not manifestly gauge invariant within the present bare current vertex (\ref{barevertex}), it can be demonstrated that the gauge invariance is kept within the dressed current vertex \cite{Feng10,Liu19}.

\section{Quantitative characteristics of electromagnetic response} \label{meissner-effect}

The electromagnetic response is quantitatively characterized by the experimentally measurable quantities, such as the local magnetic-field profile, the magnetic-field penetration depth, and superfluid density  \cite{Khasanov04,Suter04,Bozovic16,Brewer15,Deepwell13,Khasanov09,Broun07,Panagopoulos99,Lemberger11,Bernhard01,Jackson00,Kamal98,Lee96}. However, the results of the local magnetic-field profile, the magnetic-field penetration depth, and superfluid density calculated directly in terms of the response kernel in Eq. (\ref{kernel1}) can not be used for a direct comparison with the corresponding experimental results of cuprate superconductors, since this response kernel (\ref{kernel1}) evaluated based on the linear response approximation describes the electromagnetic response in an {\it infinite} system, whereas in the problem of the magnetic-field penetration and the system has a surface, i.e., it occupies a half-space $x>0$. In such a problem, we need to impose a boundary condition for electrons. This can be done within the standard specular reflection model \cite{Abrikosov88,Tinkham96} with a two-dimensional geometry of the copper-oxide plane. In this paper, we therefore discuss the magnetic-field penetration depth and superfluid density in cuprate superconductors within the copper-oxide (ab) plane only. In this case, when an applied magnetic-field is perpendicular to the ab plane, the vector potential $A_{y}(x)$ can be chosen along the $y$ axis. According to the following Maxwell equation,
\begin{eqnarray}
{\rm rot}\,{\bf B}={\rm rot}\,{\rm rot}\,{\bf A}={\rm grad}\,{\rm div}\,{\bf A}-\nabla^{2}\,{\bf A}=\mu_{0}{\bf J},
\end{eqnarray}
the extension of the vector potential in an even manner through the boundary implies a kink in the $A_{y}(x)$ curve. This reflects a fact when the magnetic-field ${\bf B}$ is applied at the system surface, i.e., $({\rm d}A_{y}(x)/{\rm d}x)|_{x=+0}=B$, while $({\rm d}A_{y}(x)/{\rm d}x)|_{x=-0} =-B$, which leading to that the second derivative $({\rm d}^{2}A_{y}(x)/ {\rm d}^{2}x)$ acquires a correction $2B\delta(x)$ \cite{Abrikosov88},
\begin{eqnarray}\label{correction}
{{\rm d}^{2}A_{y}(x)\over {\rm d}^{2}x}=2B\delta(x)-\mu_{0}J_{y},
\end{eqnarray}
where we have used the transverse gauge ${\rm div}\,{\bf A}=0$. In particular, this equation (\ref{correction}) can be Fourier transformed into the momentum space as,
\begin{eqnarray}\label{FT-correction}
q_{x}^{2}A_{y}({\bf q})=\mu_{0}J_{y}({\bf q})-2B.
\end{eqnarray}
Substituting this Fourier transform form (\ref{FT-correction}) into Eq. (\ref{linres}), and solving for the vector potential, we therefore obtain the following relation between the vector potential and the response kernel,
\begin{equation}\label{aspec}
A_{y}({\bf q})=-2B{\delta(q_{y})\delta(q_{z})\over\mu_{0}K_{yy}({\bf q})+q_{x}^{2}}.
\end{equation}
Since the vector potential has only the $y$ component, the non-zero component of the local magnetic field ${\bf h}=\rm{rot}\,{\bf A}$ is that along the $z$ axis as $h_{z}({\bf q})=iq_{x}A_{y}({\bf q})$.

\subsection{Local magnetic-field profile}\label{profile}

\begin{figure}[h!]
\centering
\includegraphics[scale=1.10]{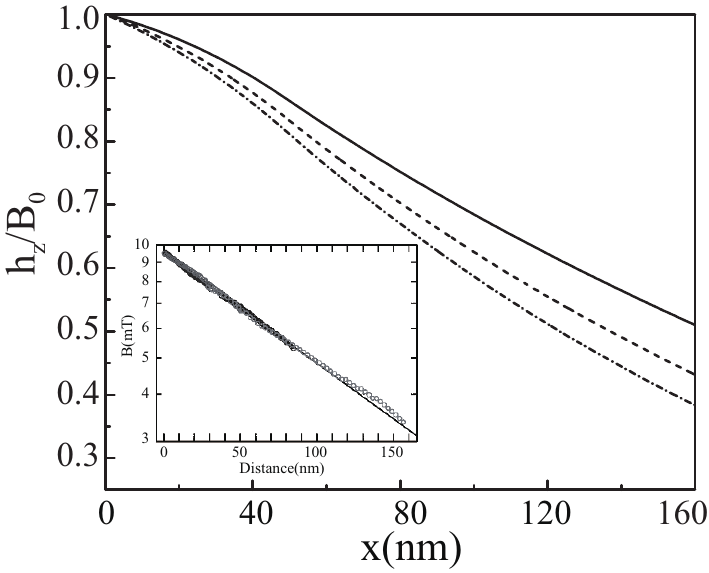}
\caption{The local magnetic-field profile as a function of the distance from the surface at $T=0.002J$ for $\delta=0.08$ (solid line), $\delta=0.12$ (dashed line), and $\delta=0.15$ (dash-dotted line) with $t/J=3.4$ and $t'/J=1.2$. Inset: the corresponding experimental result of YBa$_2$Cu$_3$O$_{7-y}$ taken from Ref. \onlinecite{Suter04}. \label{profilefig}}
\end{figure}

Now we can derive the local magnetic field profile from Eq. (\ref{aspec}) as,
\begin{equation}\label{profile}
h_{z}(x)={B\over\pi}\int\limits_{-\infty}^\infty {\rm{d}}q_{x}\,{q_{x}\sin(q_{x}x)\over\mu_{0}K_{yy}(q_{x},0,0)+q_{x}^{2}},
\end{equation}
which therefore reflects the measurably electromagnetic response in cuprate superconductors. For a convenience in the following discussions, we introduce a characteristic-length scale $a_{0}=\sqrt{\hbar^{2}a/\mu_{0}e^{2}J}$, where $a$ is the lattice constant. Using the lattice constant
$a\approx 0.383$ nm of YBa$_2$Cu$_3$O$_{7-y}$, this characteristic-length is obtained as $a_{0}\approx 97.8$ nm. In Fig. \ref{profilefig}, we plot the local magnetic-field profile $h_{z}(x)$ as a function of the distance from the surface at the temperature $T=0.002J$ for $\delta=0.08$ (solid line), $\delta=0.12$ (dashed line), and $\delta=0.15$ (dash-dotted line). For comparison, the corresponding experimental result \cite{Suter04} of YBa$_2$Cu$_3$O$_{7-y}$ is also shown in Fig. \ref{profilefig} (inset).  Apparently, the experimental result \cite{Suter04} of YBa$_2$Cu$_3$O$_{7-y}$ can be qualitatively reproduced if an external field $B=8.82$ mT is chosen to apply to the system just as it has been done in the experiment \cite{Suter04}. Moreover, the electrodynamic response is an exponential field dependence, which is also consistent with the experimental data of cuprate superconductors \cite{Khasanov04,Suter04}, however, is much different from that observed in conventional superconductors, where the local magnetic-field profile in the Meissner state shows a clear deviation from the exponential field decay \cite{Schrieffer83,Suter04}.

\subsection{Magnetic-field penetration depth}\label{penetrationdepth}

\begin{figure}[h!]
\centering
\includegraphics[scale=1.05]{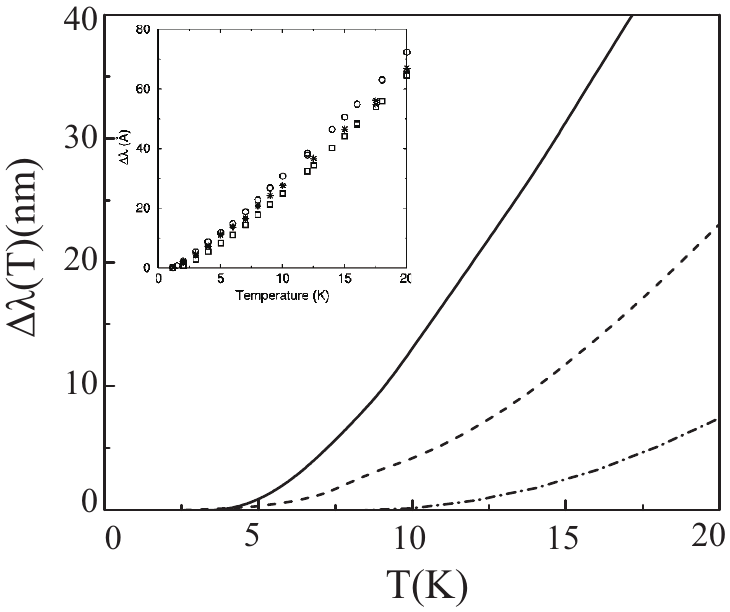}
\caption{The magnetic-field penetration depth as a function of temperature at $\delta=0.08$ (solid line), $\delta=0.12$ (dashed line), and $\delta=0.15$ (dash-dotted line) for $t/J=3.4$, $t'/J=1.2$, and $J=1000$ K. Inset: the corresponding experimental data of YBa$_2$Cu$_3$O$_{7-y}$ taken from Ref. \onlinecite{Kamal98}.\label{lambdafig}}
\end{figure}

When a cuprate superconductor is placed in an external magnetic-field $B$ smaller than the upper critical-field $B_{\rm c}$, the magnetic-field $B$ penetrates only to a penetration depth $\lambda(T)$ and is excluded from the main body of the system. This magnetic-field penetration depth $\lambda(T)$ is a fundamental parameter of cuprate superconductors, and can be evaluated directly from the local magnetic-field profile (\ref{profile}) as,
\begin{eqnarray}\label{lambda}
\lambda(T)={1\over B}\int\limits_{0}^{\infty}h_{z}(x)\,{\rm d}x={2\over\pi}\int\limits_{0}^{\infty}{\rm{d}q_{x}\over\mu_{0}K_{yy}(q_{x},0,0) +q_{x}^{2}}.~~~
\end{eqnarray}
At the temperature $T=0$, the calculated magnetic-field penetration depths are $\lambda(0)\approx 232.7$ nm, $\lambda(0)\approx 197.0$ nm, and $\lambda(0)\approx 178.5$ nm for the doping concentrations $\delta=0.08$, $\delta=0.12$, and $\delta=0.15$, respectively, which are consistent with the values of the magnetic-field penetration depth $\lambda\approx 156$ nm $\sim 400$ nm measured from different families of cuprate superconductors for different doping concentrations \cite{Khasanov04,Suter04,Bozovic16,Brewer15,Deepwell13,Khasanov09,Broun07,Panagopoulos99,Lemberger11,Bernhard01,Jackson00,Kamal98,Lee96}. On the other hand, as we have shown in subsection \ref{long-wavelength-limit}, the kernel of the response function $K_{\mu\nu}({\bf q}\to 0,0)|_{T=T{\rm c}} =0$, which leads to that $\lambda(T=T_{\rm c})=\infty$, indicating a fact that in the normal state ($T> T_{\rm c}$), the external magnetic-field can penetrate through the main body of the system, therefore there is no the Meissner effect in the normal state. To show the temperature dependence of $\lambda(T)$ more clearly, we have carried out a calculation of the evolution of $\lambda(T)$ with temperature, and the results of $\Delta\lambda(T)=\lambda(T)-\lambda(0)$ as a function of temperature at $\delta=0.08$ (solid line), $\delta=0.12$ (dashed line), and $\delta=0.15$ (dash-dotted line) are plotted in Fig. \ref{lambdafig} in comparison with the corresponding experimental results \cite{Kamal98} of YBa$_2$Cu$_3$O$_{7-y}$ (inset), where the trace of $\lambda(T)$ evolves nonlinear with temperatures at the lowest temperatures, followed by a linear temperature dependence at the low temperatures, in qualitative agreement with experimental results of cuprate superconductors \cite{Hardy93,Jackson00,Kamal98,Lee96}. However, it should be emphasized that the present result is also much different from that in conventional superconductors, where the characteristic feature is the existence of the isotropic SC gap $\bar{\Delta}$, and then $\lambda(T)$ exhibits an exponential behavior as $\lambda(T)\propto {\rm exp} (-\bar{\Delta}/T)$.

\subsection{Superfluid density}

\begin{figure}[h!]
\centering
\includegraphics[scale=1.0]{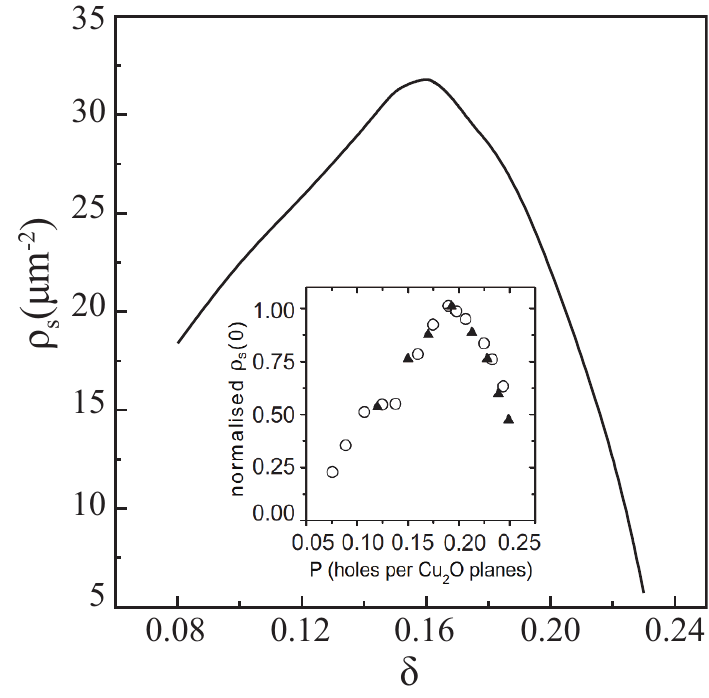}
\caption{The superfluid density as a function of doping with $T=0.002J$ for $t/J=3.4$ and $t'/J=1.2$. Inset: the corresponding experimental result of Y$_{0.8}$Ca$_{0.2}$Ba$_{2}$(Cu$_{1-z}$Zn$_{z}$)$_{3}$O$_{7-\delta}$ (open circles) and Tl$_{1-y}$Pb$_{y}$Sr$_{2}$Ca$_{1-x}$Y$_x$Cu$_2$O$_{7}$ (solid triangles) taken from Ref. \onlinecite{Bernhard01}. \label{rhofig}}
\end{figure}

The superfluid density $\rho_{\rm s}(T)$ plays a special role in the physics of cuprate superconductors, as it describes the SC quasiparticles and determines the stiffness of the SC order parameter. This $\rho_{\rm s}(T)$ is obtained from the inverse square of the magnetic-field penetration depth as,
\begin{eqnarray}
\rho_{\rm s}(T)={1\over \lambda^{2}(T)}. \label{rhodensity}
\end{eqnarray}
As a natural consequence of doped Mott insulators, this $\rho_{\rm s}$ is strongly dependent on doping. To show this point clearly, we plot $\rho_{\rm s}$ as a function of doping with $T=0.002J$ in Fig. \ref{rhofig}. For comparison, the corresponding experimental result \cite{Bernhard01} of Y$_{0.8}$Ca$_{0.2}$Ba$_{2}$(Cu$_{1-z}$Zn$_{z}$)$_{3}$O$_{7-\delta}$ and Tl$_{1-y}$Pb$_{y}$Sr$_{2}$Ca$_{1-x}$Y$_x$Cu$_2$O$_{7}$ is also shown in Fig. \ref{rhofig} (inset). Our result in Fig. \ref{rhofig} shown clearly that in a striking analogous to the domelike shape doping dependence $T_{\rm c}$, the superfluid density $\rho_{\rm s}$ appears from the starting point of the SC dome, and then increases with the increase of doping in the lower doped regime, however, this $\rho_{\rm s}$ reaches its highest value at around the {\it critical doping} $\delta_{\rm critical}\approx 0.16$, and then decreases with the increase of doping at the higher doped regime, eventually disappearing together with superconductivity at the end of the SC dome. In particular, this anticipated critical doping $\delta_{\rm critical}\approx 0.16$ for the highest $\rho_{\rm s}$ is not too far from the critical doping $\sim 0.19$ for the highest $\rho_{\rm s}$ estimated from cuprate superconductors \cite{Lemberger11,Bernhard01}. This doping dependence of $\rho_{\rm s}$ therefore also indicates that the close correlation between $T_{\rm c}$ and $\rho_{\rm s}$ is a significant and intrinsic feature of cuprate superconductors.

\begin{figure}[h!]
\centering
\includegraphics[scale=1.10]{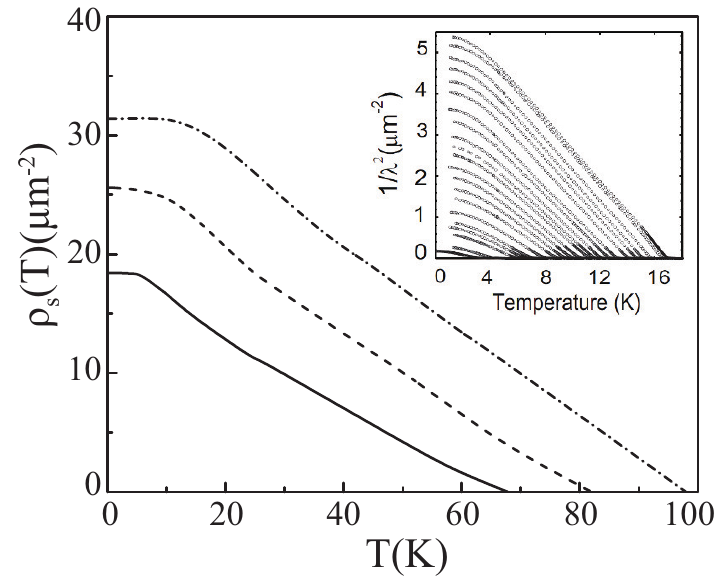}
\caption{The superfluid density as a function of temperature at $\delta=0.08$ (solid line), $\delta=0.12$ (dashed line), and $\delta=0.15$ (dash-dotted line) for $t/J=3.4$, $t'/J=1.2$, and $J=1000$ K. Inset: the corresponding experimental result of YBa$_2$Cu$_3$O$_{7-y}$ taken from Ref. \onlinecite{Broun07}. \label{rhofigt}}
\end{figure}

The evolution of $\lambda(T)$ with temperature shown in Fig. \ref{lambdafig} also induces the temperature dependence of $\rho_{\rm s}(T)$. However, as we have shown in Subsection \ref{penetrationdepth}, at the temperatures $T=T_{\rm c} $, $\lambda(T=T_{\rm c})=\infty$, which therefore leads to $\rho_{\rm s}(T=T_{\rm c})=0$, in good agreement with the result of the effective superfluid density shown in Fig. \ref{effectivsuperfluid}. For a further understanding of the unusual temperature dependence of $\rho_{\rm s}(T)$, we plot $\rho_{\rm s}(T)$ as a function of temperature at $\delta=0.08$ (solid line), $\delta=0.12$ (dotted line), and $\delta=0.15$ (dash-dotted line) in Fig. \ref{rhofigt} in comparison with the corresponding experimental result \cite{Broun07} of YBa$_2$Cu$_3$O$_{7-y}$ (inset), where in the corresponding to the nonlinear temperature dependence of $\lambda(T)$ at the lowest temperatures shown in Fig. \ref{lambdafig}, $\rho_{\rm s}(T)$ shows a nonlinear temperature behavior at the lowest temperatures. However, the most striking feature is the strong linear temperature dependence extending over the most of the SC range, which is, of course, the expected one for the kinetic-energy driven superconductivity with the d-wave symmetry. Incorporating the results in Fig. \ref{rhofig} and Fig. \ref{rhofigt}, it is thus shown that $\rho_{\rm s}(T)$ is both correlated with  $T_{\rm c}$ and the linear in temperatures. These theoretical results are also in qualitative agreement with the experimental data \cite{Bozovic16,Brewer15,Deepwell13,Khasanov09,Broun07,Panagopoulos99} of cuprate superconductors.

An explanation of the nonlinearity in the temperature dependence of $\rho_{\rm s}(T)$ in cuprate superconductors at the lowest temperatures can be found from the nonlocal effect induced by the d-wave gap nodes on EFS \cite{Feng10,Kosztin97,Yip92}. This follows from a fact that from the general relation in Eq. (\ref{linres}), the nonlocal relation between the supercurrent and vector potential in the coordinate space is valid due to the finite size of the electron pairs. Moreover, although the external magnetic-field decays exponentially on the scale of the magnetic-field penetration depth $\lambda(T)$, any nonlocal contributions to the measurable quantities are of the order of $\kappa^{-2}$, where the Ginzburg--Landau parameter $\kappa$ is the ratio of $\lambda$ and the coherence length $\zeta$. In the present kinetic-energy driven superconductivity with the d-wave symmetry \cite{Feng15,Feng0306,Feng12}, the SC gap $\bar{\Delta}({\bf k})$ is correlated with the coherence length $\zeta({\bf k})$ in terms of the equation $\zeta({\bf k})=\hbar v_{\rm F}/[\pi\bar{\Delta}({\bf k})]$, where $v_{\rm F}=\hbar^{-1}\partial\varepsilon_{\bf k}/\partial {\bf k}|_{k_{\rm F}}$ is the electron velocity at EFS. In particular, the size of the electron pairs in the clean limit is of the order of this coherence length $\zeta({\bf k})$, and therefore is momentum dependent. However, for cuprate superconductors, the d-wave SC gap disappears at around the gap nodes on EFS, leading to the existence of the gapless quasiparticle excitations. These gapless quasiparticle excitations induces a divergence of the coherence length $\zeta({\bf k})$ at around the gap nodes on EFS, and then the behavior of the temperature dependence of $\rho_{\rm s}(T)$ depends sensitively on the quasiparticle scattering at around the gap nodes. At the lowest temperatures, the most of the quasiparticles are accommodated selectively at around the gap nodes on EFS, and therefore the main contribution to the measurable quantities comes from these quasiparticles. In this case, the Ginzburg--Landau ratio $\kappa({\bf k})$ at around the gap nodes on EFS is no longer large enough for the system to belong to the class of type-II superconductors, and the condition of the local limit is not fulfilled \cite{Kosztin97}, which leads to the system in the extremely nonlocal limit, and then the nonlinear behavior in the temperature dependence of $\rho_{\rm s}(T)$ is observed experimentally \cite{Bonn96,Khasanov09,Broun07,Panagopoulos99}. On the other hand, with increasing temperature, the quasiparticles at around the gap nodes on EFS become excited out of the condensate, and then the nonlocal effect vanishes, where the momentum dependent coherence length $\zeta({\bf k})$ can be replaced approximately with the isotropic one $\zeta_{0}=\hbar v_{\rm F}/[\pi\bar{\Delta}]$, where $\bar{\Delta}$ is the d-wave gap amplitude at zero temperature. In this case, the calculated Ginzburg--Landau parameters are $\kappa_{0}\approx\lambda(0)/\zeta_{0}\approx 7.86$, $\kappa_{0}\approx 13.05$, and $\kappa_{0}\approx 18.20$ for the doping concentrations $\delta=0.08$, $\delta=0.12$, and $\delta=0.15$, respectively, and then the condition of the local limit is satisfied. In particular, these theoretical values of the Ginzburg--Landau parameter at different doping concentrations match well with the experimental estimations for different families of cuprate superconductors at different doping concentrations \cite{Khasanov04,Suter04,Bozovic16,Brewer15,Deepwell13,Khasanov09,Broun07,Panagopoulos99,Lemberger11,Bernhard01,Jackson00,Kamal98,Lee96}. As a consequence, the present results show that cuprate superconductors at the low temperatures turn out to be type-II superconductors, where nonlocal effects can be neglected, then the electrodynamics is purely local and the magnetic-field decays exponentially over a length of the order of a few hundreds nm.

\section{Conclusions}\label{conclusions}

Within the kinetic-energy driven SC mechanism, we have discussed the doping and temperature dependence of the electromagnetic response in cuprate superconductors. We evaluate the kernel of the response function based on the linear response approximation for a purely transverse vector potential, and find that it can be separated into two parts: one associated with the diamagnetic current, while the other with the paramagnetic current. In particular, this paramagnetic part of the response kernel exactly cancels the corresponding diamagnetic part of the response kernel in the normal-state, and then the Meissner effect is obtained within the entire SC phase. Following this kernel of the response function, the electromagnetic response calculation in terms of the specular reflection model qualitatively reproduces many of the striking features observed in the experiments. In particular, the local magnetic-field profile follows an exponential law, while the superfluid density exhibits the nonlinear temperature behavior at the lowest temperatures, followed by the linear temperature dependence extending over the most of the SC temperature range. Moreover, the superfluid density takes a domelike shape with the lower-doped and higher-doped regimes on each side of the {\it critical doping} $\delta_{\rm critical}\approx 0.16$, where the magnitude of the superfluid density reaches its maximum. Our theory also shows that the nonlinear temperature dependence of the superfluid density at the lowest temperatures is attributed to the nonlocal effects induced by the d-wave gap nodes on EFS.

Finally, it should be emphasized that in the present study based on the kinetic-energy driven superconductivity, the main purpose is to illustrates the qualitative features of the electromagnetic response contained within the clean-limit d-wave superconductivity. However, the calculation performed in the present study does not include the pseudogap effect, which is known to be very important for cuprate superconductors. In our recent studies, we have shown that EFS in cuprtae superconductors is reconstructed due to presence of the pseudogap \cite{Feng16}. In particular, this EFS instability drives the charge-order correlation \cite{Feng16,Gao18}, generating a coexistence of superconductivity and charge order. In this state of the coexistence of superconductivity and charge order, we \cite{Liu19} have also discussed the temperature dependence of the superfluid density by taking into account the interplay between the SC gap and pseudogap, and the obtained results show that although the global feature of the superfluid density does not change so much, the magnitude of the superfluid density in the underdoped regime is suppressed by the pseudogap. These and the related results will be presented elsewhere.


\section*{Acknowledgements}

This work was supported by the National Key Research and Development Program of China under Grant No. 2016YFA0300304, and the National Natural Science Foundation of China under Grant Nos. 11574032 and 11734002.

\end{document}